\documentclass[aps, prd, twocolumn, amsmath, floats,floatfix, superscriptaddress, nofootinbib,
showpacs]{revtex4}

\usepackage{amssymb}
\usepackage{amsmath}
\usepackage{verbatim}
\usepackage{mathrsfs}
\usepackage{amsfonts}
\usepackage{latexsym}
\usepackage{epsfig}
\usepackage{color}
\usepackage{graphicx,subfigure}

\begin{document}

\title{Quasistationary solutions of scalar fields around accreting black holes} 

\author{Nicolas Sanchis-Gual}
\affiliation{Departamento de
  Astronom\'{\i}a y Astrof\'{\i}sica, Universitat de Val\`encia,
  Dr. Moliner 50, 46100, Burjassot (Val\`encia), Spain}

\author{Juan Carlos Degollado} 
\affiliation{Instituto de Ciencias F\'isicas, Universidad Nacional Aut\'onoma de M\'exico,
Apdo. Postal 48-3, 62251, Cuernavaca, Morelos, M\'exico.}

\author{Paula Izquierdo}
\affiliation{Universidad de la Laguna (ULL), Av. Astrof\'{\i}sico Francisco S\'{a}nchez S/N, 38206, 
San Crist\'{o}bal de La Laguna, Santa Cruz de Tenerife, Spain}  

\author{Jos\'e A. Font}
\affiliation{Departamento de
  Astronom\'{\i}a y Astrof\'{\i}sica, Universitat de Val\`encia,
  Dr. Moliner 50, 46100, Burjassot (Val\`encia), Spain}
\affiliation{Observatori Astron\`omic, Universitat de Val\`encia, C/ Catedr\'atico 
  Jos\'e Beltr\'an 2, 46980, Paterna (Val\`encia), Spain}
  
\author{Pedro J. Montero} 
\affiliation{Max-Planck-Institute f{\"u}r Astrophysik, Karl-Schwarzschild-Str. 1, 85748, Garching bei M{\"u}nchen, Germany}


\today


\begin{abstract}  
Massive scalar fields can form long-lived configurations around black holes. These 
configurations, dubbed quasi-bound states, have been studied both in the linear and nonlinear regimes.
In this paper we show that quasi-bound states can form in a dynamical scenario in which the mass of the black hole 
grows significantly due to the capture of infalling matter. We solve the Klein-Gordon equation numerically in 
spherical symmetry, mimicking the evolution of the spacetime through a sequence of analytic Schwarzschild 
black hole solutions of increasing mass. It is found that the frequency of oscillation of the quasi-bound 
states decreases as the mass of the black hole increases. In addition, accretion 
leads to a significative increase of the exponential decay of the scalar field energy due to the presence of terms of 
order higher than linear in the exponent. We compare the black hole mass growth rates used in our study 
with estimates from observational surveys and extrapolate our results to values 
of the scalar field masses consistent with models that propose scalar fields as dark matter in the universe. 
We show that even for unrealistically large mass accretion rates, quasi-bound states around 
accreting black holes can survive for cosmological timescales. Our results provide further support to the intriguing 
possibility of the existence of dark matter halos based on (ultra-light) scalar fields surrounding supermassive 
black holes in galactic centers.
\end{abstract}




\maketitle

\section{Introduction}\label{sec:introduction}

There is compelling evidence that most nearby galaxies host supermassive black holes (SMBHs) in their centers, with 
masses in the range $10^6-10^9 M_{\odot}$~\cite{Ferrarese2005,Kormendy2013}. Such SMBHs  are expected 
to be the evolutionary result of the growth of seed BHs in high redshift galaxies through accretion episodes and
mergers of massive BHs (MBH) binaries with masses somewhere in between those of stellar-origin BHs and 
SMBHs~\cite{Kauffmann2000,Volonteri2003}. The discovery of supermassive luminous quasars 
at redshifts of $z > 6$ has shown that SMBHs with masses $\sim 10^9 M_{\odot}$, must form 
extremely early on in the history of the universe~\cite{Barth2003,Willott2005,mortlock2011luminous}. These SMBHs
must grow rapidly in order to acquire its mass within a short period  of $\sim 1$ Gyr.
Explaining the formation and evolution of SMBHs dwelling in the most powerful quasars when the
Universe  was  less  than  1 Gyr  old  (and  of  the regular and much  smaller  MBHs  hidden  in  13 Gyr  old
galaxies) is a pressing open issue \cite{volonteri2010formation}. Proposed models involve  the 
gravitational collapse of gas clouds or the collapse of supermassive stars in the early universe (a model hampered by the
low masses of the initial seeds of first-generation Pop III stars \cite{madau2001massive}), the runaway growth 
by accretion onto Pop III BH,  or mergers of smaller size BHs. In either scenario, the formation of SMBHs is a 
highly dynamical event amenable to gravitational wave investigations. Indeed, it is expected that eLISA will 
probe MBH binaries  in  the  $10^3-10^7 M_{\odot}$ range  out  to redshift $z>10$ through the detection of their 
gravitational waves in the mHz frequency band~\cite{Barausse2015}.

While one can make a convincing case that SMBHs have grown largely through accretion, with the
consequent energy emission observed in electromagnetic output, it
has been argued that an exponential growth at the Eddington-limited e-folding time is too slow to 
grow stellar-mass BH seeds into the supermassive luminous quasars that are observed at $z \sim 7$ \cite{mortlock2011luminous}.
Some proposals to circumvent this issue invoke super-Eddington accretion for brief 
periods of time \cite{alexander2014rapid}, the formation through accretion and merging of 
the first stellar remnants \cite{bromm2003formation,haiman2001highest} or via more
massive seeds from the collapse of pre-galactic disks at high redshifts \cite{begelman2006formation,begelman2010evolution,mayer2010direct}.

In light of these proposals it is worth considering if rapid BH accretion may have any effect 
on the distribution of the associated dark matter halo when the latter is modeled as a scalar field.
In the absence of accretion the existence of long-lasting scalar field configurations surrounding a
(non-rotating) BH has been investigated in a number of recent papers, either in the test-field
limit~\cite{Barranco2011,Barranco:2012qs,Barranco:2013rua}  or employing self-gravitating 
scalar fields~\cite{Sanchis-Gual:2015bh,Sanchis-Gual:2015sms}.  
These papers have shown that SMBHs at galactic centers do not
represent a serious threat to dark matter models based on (ultra-light) scalar fields as a viable 
alternative to the usual description of dark matter 
in terms of weakly interacting massive particles. For both, scalar fields around SMBHs and
axion-like scalar fields around primordial BHs, it has been found that scalar fields can survive for cosmological
timescales~\cite{Barranco:2012qs}.
Despite these findings, and due to the rapid growth of accretion-powered SMBHs, it is worth investigating the chances of 
survival of the scalar field within such dynamical situation. This is the purpose of this paper. Here
we study the properties of scalar field quasi-bound states in the background of an accreting spherically symmetric BH.
Assuming that the BH mass grows adiabatically due to infalling matter we show 
that indeed, long-lasting, quasi-bound states can still form in such scenario. 

This paper is organized as follows: in Section~\ref{sec:formalism} we lay out the mathematical and physical approach
we use to carry out our investigation. In particular Section~\ref{sec:numerics} contains a brief synopsis of our numerical 
methodology. The results are presented and discussed in Section~\ref{sec:results}, while Section~\ref{sec:summary} 
contains the summary of our findings. Throughout the paper we employ geometrized units ($c=G=1$). Latin indices indicate 
spatial indices and hence run from 1 to 3 while Greek indices run from 0 to 3.
 
\section{Problem setup}\label{sec:formalism}

\subsection{Klein-Gordon equation}

Our setup considers a scalar field distribution $\Phi$ of sufficiently small energy to neglect its self-gravity, i.e.~the field
can be regarded as a
test-field. This configuration surrounds a BH which is assumed to be continuously accreting matter. The dynamics of the
scalar field is governed by the Klein-Gordon equation,
\begin{equation}
 \Box \Phi-\mu^2\Phi=0 \ ,
\label{eq:KG}
\end{equation}
where the D'Alambertian operator is defined by $\Box:=
(1/\sqrt{-g})\partial_{\alpha}(\sqrt{-g}g^{\alpha\beta}\partial_{\beta})$. We
follow the convention that $\Phi$ is dimensionless and $\mu$, the mass of the scalar field, has
dimensions of (length)$^{-1}$. 

We write the spacetime metric $g_{\alpha\beta}$ as
\begin{eqnarray} \label{metric}
ds^2 & = & g_{\alpha\beta} dx^\alpha dx^\beta \nonumber \\ 
& = & - \alpha^2 dt^2 + \gamma_{ij} (dx^i + \beta^i dt)(dx^j + \beta^j dt),
\end{eqnarray}
where $\alpha$ is the lapse function, $\beta^i$ the shift vector, and $\gamma_{ij}$ the spatial metric. We adopt a conformal decomposition of the spatial metric $\gamma_{ij}$ 
\begin{equation} \label{conformal_decomposition}
\gamma_{ij} = e^{4 \chi} \hat \gamma_{ij},
\end{equation}
where $\psi := e^\chi$ is the conformal
factor, $\hat \gamma_{ij}$ the conformally related metric and $\hat
\gamma$ its determinant. By assuming spherical symmetry the line element may be
written as
\begin{equation}
 dl^2 = e^{4\chi } (a(t,r)dr^2+ r^2\,b(t,r)  d\Omega^2)\,,
\end{equation}
with $d\Omega^2 = \sin^2\theta d\varphi^2+d\theta^2$ being the solid angle element
and $a(t,r)$ and $b(t,r)$ two independent metric functions.

To solve the Klein-Gordon equation in spherical symmetry, we introduce two first-order fields, $\Pi$ and $\Psi$,  
defined as:
\begin{eqnarray}
\Pi &:=& n^{\alpha}\partial_{\alpha}\Phi=\frac{1}{\alpha}(\partial_{t}\Phi-\beta^{r}\partial_{r}\Phi) \ ,\\
\Psi&:=&\partial_{r}\Phi \ ,
\end{eqnarray}
where $n^{\alpha}$ is the unit vector normal to the surfaces of constant $t$.
Therefore, using Eq.~(\ref{eq:KG}) we obtain the following system of first-order equations: 
\begin{eqnarray}
\partial_{t}\Phi&=&\beta^{r}\partial_{r}\Phi+\alpha\Pi \,,\\
\partial_{t}\Psi&=&\beta^{r}\partial_{r}\Psi+\Psi\partial_{r}\beta^{r}+\partial_{r}(\alpha\Pi) \,,\\
\partial_{t}\Pi&=&\beta^{r}\partial_{r}\Pi+\frac{\alpha}{ae^{4\chi}}[\partial_{r}\Psi\nonumber\\
&+&\Psi\biggl(\frac{2}{r}-\frac{\partial_{r}a}{2a}+\frac{\partial{r}b}{b}+2\partial_{r}\chi\biggl)\biggl]\nonumber\\
&+&\frac{\Psi}{ae^{4\chi}}\partial_{r}\alpha+\alpha K\Pi - \alpha \mu^{2}\Phi \,,
\label{eq:sist-KG}
\end{eqnarray}
where $K$ is the trace of the extrinsic curvature.
The stress-energy tensor of the scalar field reads
\begin{equation}
T_{\alpha\beta}=\partial_{\alpha}\Phi \partial_{\beta}\Phi-\frac{1}{2}g_{\alpha\beta}\left(\partial^{\sigma}\Phi\partial_{\sigma}\Phi+
\mu^2\Phi^2 \right)\,.
\label{eq:tmunu}
\end{equation}
From this tensor we can compute the energy of the scalar field $E$
\begin{equation}\label{eq:scalar}
E_{\text{SF}}=\int_{r_{\rm{AH}}}^{\infty}\mathcal{E}_{\text{SF}} dV  \,,
\end{equation}
where $r_{\rm{AH}}$ is the radius of the apparent horizon and $\mathcal{E}^{\text{SF}}$ is given by 
\begin{eqnarray}
\mathcal{E}_{\text{SF}}&:=&n^{\alpha}n^{\beta}T_{\alpha\beta}=\frac{1}{2}\biggl(\Pi^{2}+\frac{\Psi^{2}}{ae^{4\chi}}\biggl)+\frac{1}{2}\mu^{2}\Phi^{2} \label{eq:rho}\,.
\end{eqnarray}
%

\subsection{Analytic black hole solution}
\label{analytic}

The numerical evolution of a single BH using the so-called ``moving puncture" technique leads to a 
well-known time-independent, maximal slicing solution of Schwarzschild~\cite{Hannam:2007prl}. 
It was shown in \cite{Baumgarte:2007a} that this solution can also be constructed analytically and be 
used as a test for numerical codes. We take advantage of this result to put forward the defining 
characteristic of the procedure we employ in the current investigation. Namely, we avoid evolving 
numerically the BH, computing instead a sequence of analytical solutions at each time step for 
different BH masses to mimic the BH growth for different accretion rates. This procedure allows us 
to simulate long episodes of accretion without resorting to a test fluid in order to achieve high 
accretion rates.

The analytic solution is constructed by defining all quantities as a function of the gauge-invariant areal radius $R$. We have to transform the solution into isotropic coordinates, as the latter are used in our numerical procedure, comparing the spatial metrics as a function of $R$ and the isotropic radius $r$
\begin{eqnarray}
\alpha^{-2}\,dR^{2}+R^{2}d\Omega^{2}=\psi^{4}(dr^{2}+r^{2}d\Omega^{2})\,.
\end{eqnarray}
Thus, the isotropic radius $r$ is given as a function of $R$ by

\begin{eqnarray}\label{eq:riso}
r&=&\biggl[\frac{\displaystyle 2R+M+(4R^{2}+4MR+3M^{2})^{1/2}}{4}\biggl]\nonumber\\
&\times&\biggl[\frac{4+3\sqrt{2})(2R-3M)}{8R+6M+3(8R^{2}+8MR+6M^{2})^{1/2}}\biggl]^{1/\sqrt{2}}\nonumber\\
&=&R\biggl[1-\frac{M}{R}-\frac{M^{2}}{2R^{2}}+...\biggl].
\end{eqnarray}
In the limiting case $R\rightarrow3M/2$, $r\rightarrow0$. Correspondingly, the conformal factor is obtained from
\begin{eqnarray}\label{eq:conformaliso}
\psi&=&\biggl[\frac{4R}{\displaystyle 2R+M+(4R^{2}+4MR+3M^{2})^{1/2}}\biggl]\nonumber\\
&\times&\biggl[\frac{8R+6M+3(8R^{2}+8MR+6M^{2})^{1/2}}{4+3\sqrt{2})(2R-3M)}\biggl]^{1/2\sqrt{2}}\,.
\end{eqnarray}

Finally, the lapse function and the isotropic shift vector are respectively given by
\begin{eqnarray}\label{eq:alpiso}
\alpha&=&\biggl(1-\frac{2M}{R}+\frac{27M^{4}}{16R^{4}}\biggl)^{1/2}\,,
\\
\beta^{r}&=&\frac{3\sqrt{3}M^{2}}{4}\frac{r}{R^{3}}\,.\label{eq:betaiso}
\end{eqnarray}

\subsection{Adiabatic growth of the BH mass }

We assume that the mass of the BH grows due to the capture of matter as e.g.~falling in from an
accretion disk. This infalling matter is assumed to interact with the scalar field only gravitationally,
that is the quasi-bound states surrounding the BH can only be affected by the increase of the mass 
of the BH. We will employ a simple phenomenological law based on observational and theoretical 
grounds  which allows us to reasonably incorporate the growth of the BH mass in our model.
As mentioned before, the observational evidence of the existence of very luminous quasars, which 
implies BH masses of $\sim10^{8}-10^{9}M_{\odot}$ at 
$z\sim6-7$~\cite{willott2010eddington,fan2006constraining,mortlock2011luminous} demonstrates 
that SMBHs grow rapidly in a short span of time ($\sim10^{9}$ years). Moreover, cosmological 
simulations~\cite{springel2005modelling,hu2006forming,di2008direct} suggest that SMBH seeds 
undergo an exponential growth phase at early times, $z\gtrsim4$. Therefore, given a growth rate 
$\dot{M}_{\rm{BH}}$, the mass of the BH will increase as
\begin{equation}\label{eq:bhmass}
 M = M_{0}\,e^{\dot{M}_{\rm{BH}}  t \ },
\end{equation}
where $M_0$ is the initial BH mass and $t$ is the time in our simulations.
The actual mechanism that produces the growth of the BH mass is not actually relevant for our
study since we are interested in describing the evolution of the scalar field. Thus, assuming that it grows 
according to Eq.~(\ref{eq:bhmass}) seems quite convenient.

\begin{figure}
\begin{minipage}{1\linewidth}
\includegraphics[width=1.0\textwidth, height=0.3\textheight]{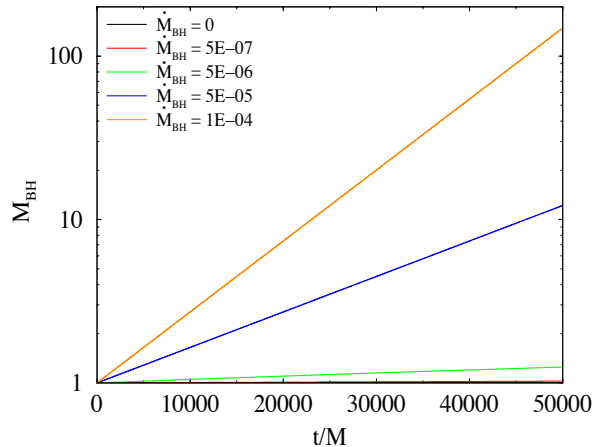} 
\caption{Evolution of the BH mass for the different accretion rates indicated in the legend. Note that the cases $\dot{M}_{\rm{BH}}=0$ and $\dot{M}_{\rm{BH}}=5\times 10^{-7}$ practically overlap in the plot. Unless stated otherwise the same color criterion
for $\dot{M}_{\rm{BH}}$ is employed in the remaining figures of this paper.}
\label{fg:BH}
\end{minipage}
\end{figure}

We first consider five different values for $\dot{M}_{\rm{BH}}$, namely, $\dot{M}_{\rm{BH}}=\lbrace0, 5\times 10^{-7}, 
5\times 10^{-6}, 5\times 10^{-5}$ and $10^{-4\rbrace}$. The time evolution of the BH mass for the different growth (accretion) rates considered is 
plotted in Fig.~\ref{fg:BH}. 
We note that our values of $\dot{M}_{\rm{BH}}$, while small, are nevertheless orders of magnitude higher than realistic values 
inferred from observations, which are at most $\dot{M}_{\rm{BH}}\sim10^{-11}$ in our units~\cite{Tsai:2015wise}. (We further
expound on this issue in Section~\ref{comparison-observations} below.) Using such values would however render the 
numerical investigation prohibitively expensive.
The first set of simulations reported in this paper evolve the scalar fields up to a final time $t=4\times 10^4M$, 
which is $\sim0.2$ s for $M=1\,M_{\odot}$. 
Despite the evolutions are fairly long from the computational point of view, they are certainly not so on astrophysical grounds. Therefore,
in order to study the effect of accretion on the scalar field evolution in affordable computational times, we have to resort to large enough growth rates. As we show below, our simulations indicate that even for the smaller $\dot{M}_{\rm{BH}}$ considered some slight differences appear by the end of the simulation. Nevertheless, longer, and computationally more expensive, simulations would be necessary to show the influence of the growth rate for the smaller values of $\dot{M}_{\rm{BH}}$.

In the second part of our study we change the computational setup of the problem by placing reflecting boundary conditions for the 
scalar field at some radius, i.e.~placing a {\it mirror} beyond which the scalar field is required to vanish, as we did 
in~\cite{Sanchis-Gual:2016}. While evolving the scalar field in a cavity is certainly an unrealistic situation, it has some practical 
advantages as it allows us to investigate lower values of $\dot{M}_{\rm{BH}}$ and perform longer runs since the computational 
grid is significantly smaller than in the ``natural" setup (with outgoing boundary conditions at the outermost radial grid zone). Within 
this idealized setup the initial value of $\dot{M}_{\rm{BH}}$ can be as low as $\dot{M}_{\rm{BH}}=5\times 10^{-9}$. 
Despite this reduction only brings in practice about one order of magnitude gain in the final evolution time, it is nevertheless 
significant to better quantify the exponential decay in the energy of the scalar field, as we show below.

\subsection{Initial data}
\label{ID}

As initial data for the scalar field we set a Gaussian distribution of the form
\begin{equation}\label{eq:pulse}
 \Phi=A_0e^{-(r-r_0)^2/\lambda^2} \ ,
\end{equation}
where $A_0$ is the initial amplitude, $r_0$ is the center of the Gaussian, and
$\lambda$ is its width. The auxiliary first order quantities are
initialized as follows
\begin{eqnarray}
 \Pi(t=0,r)&=&0 \ , \\ 
 \Psi(t=0,r) &=& -2\frac{(r-r_0)}{\lambda^2}A_0e^{-(r-r_0)^2/\lambda^2} \ .
 \label{eq:iderivatives}
\end{eqnarray}
Likewise, we choose a conformally flat metric with $a=b=1$ together with a time symmetry condition, i.e.~a vanishing extrinsic
curvature, $K_{ij}=0$. 

\begin{figure*}
\begin{center}
\subfigure{\includegraphics[width=0.45\textwidth]{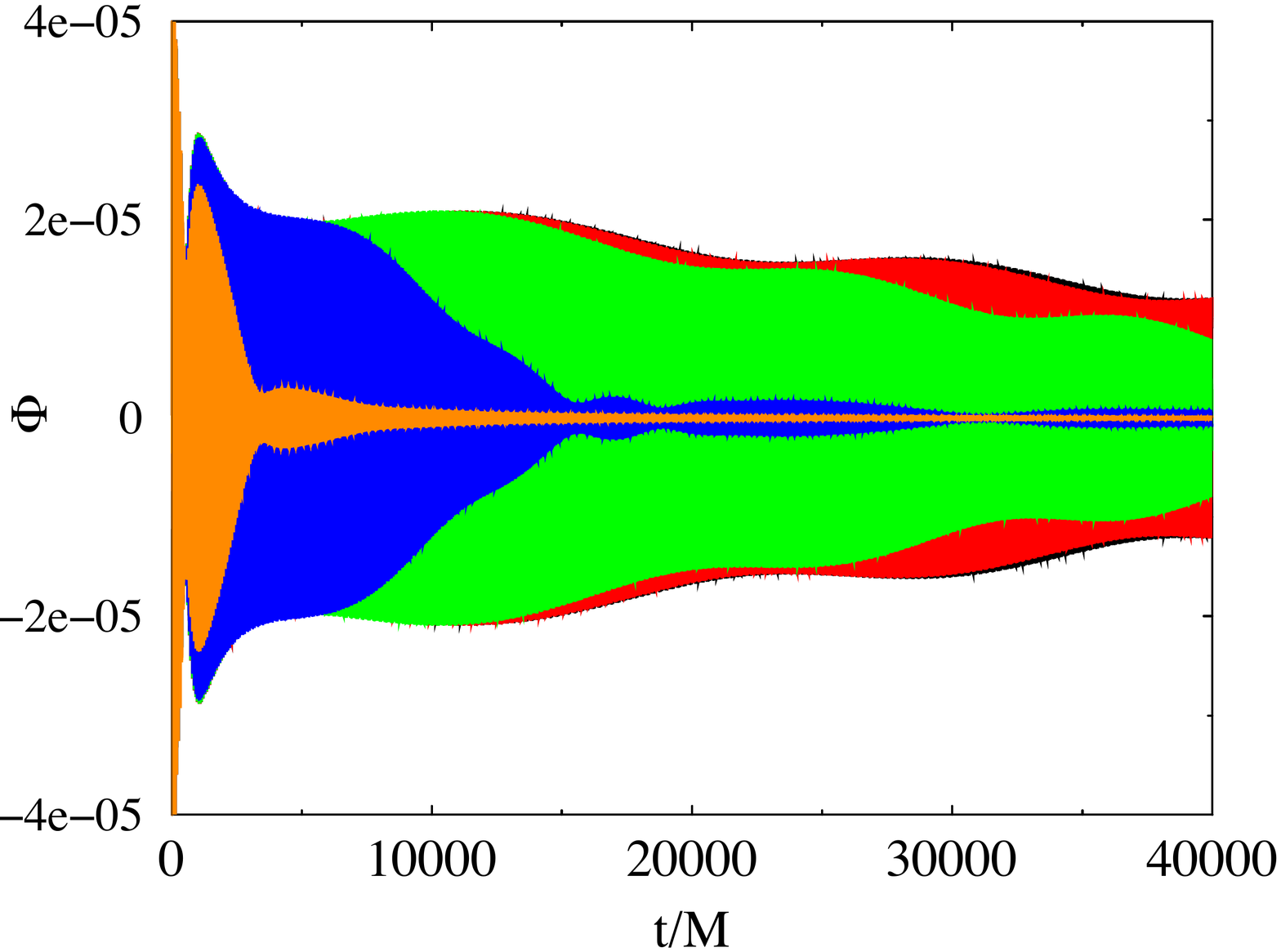}}\hspace{-0.8cm}
\subfigure{\includegraphics[width=0.45\textwidth]{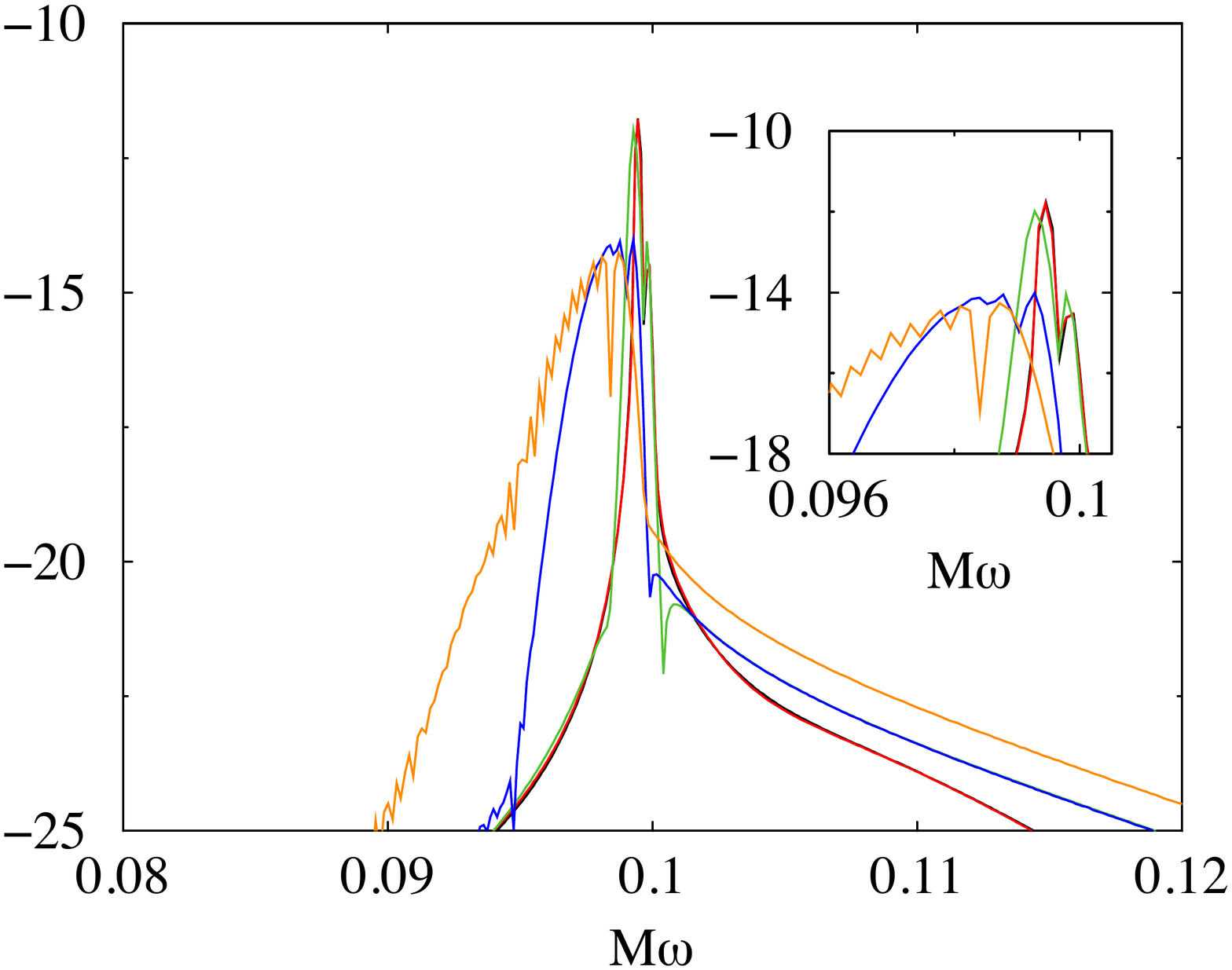}}\vspace{-1.0cm}\\
\subfigure{\includegraphics[width=0.45\textwidth]{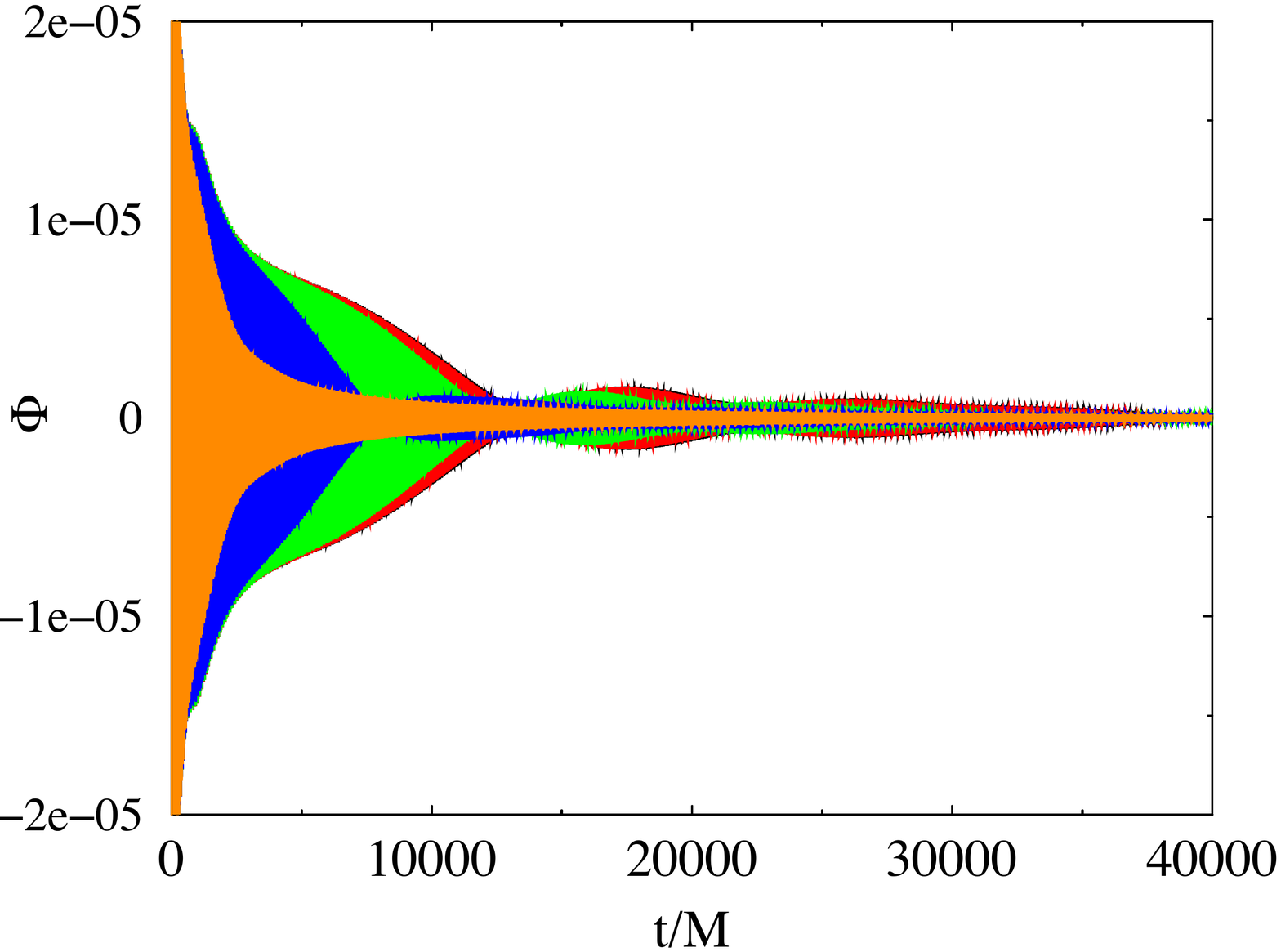}}\hspace{-0.8cm}
\subfigure{\includegraphics[width=0.45\textwidth]{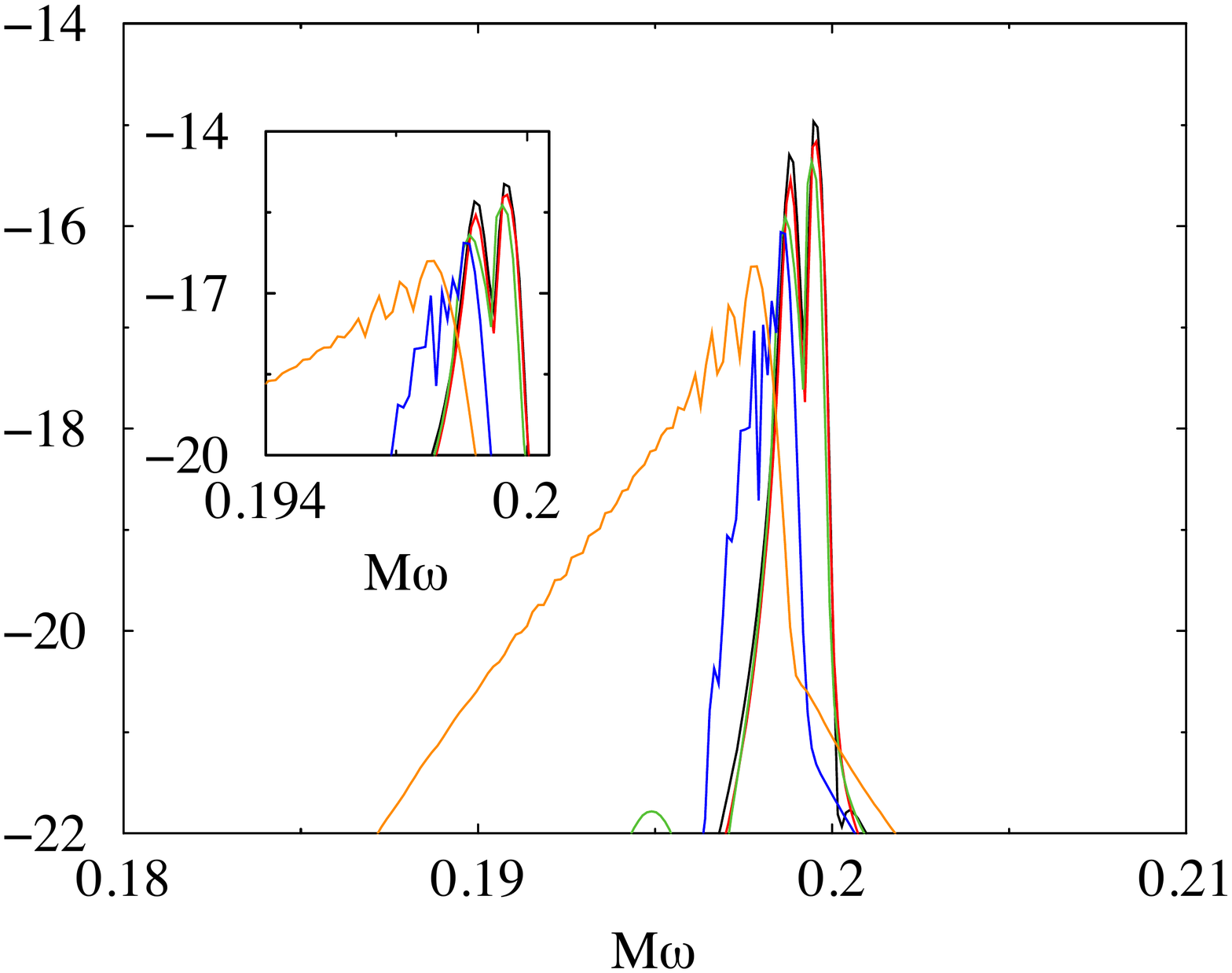}}
\caption{{\it Left column}: Time evolution of the scalar field with mass $M\mu=0.1$ (top)  and $M\mu=0.2$ (bottom) and for different BH growth rates, $\dot{M}_{\rm{BH}}$. {\it Right column}: Corresponding power spectra obtained from Fourier transforming the time series shown on the left. The units in the vertical axis are arbitrary.}
\label{fg:SF1}
\end{center}
\end{figure*}

For the BH, we compute the Schwarzschild solution using the moving puncture technique from equations~(\ref{eq:riso})-(\ref{eq:betaiso}) for the initial mass $M_{0}=1$.

\begin{table*}
\caption{Initial parameters and most relevant quantities for the different distributions of scalar fields considered. 
From left to right the columns report: the BH mass growth rate, $\dot M_{\rm{BH}}$, the scalar field mass, $M\mu$, 
the initial amplitude of the pulse, $A_0$, the real part of the angular frequency $\omega$ for the fundamental mode of 
oscillation and the first overtone, the linear, quadratic, and cubic numerical fits of the decay rate 
of the modes (coefficients $a$, $b$ and $c$ in the text), $Ms$, and the final BH mass, $M_{\rm{BH}}$. 
The initial Gaussian pulse is located at $r_0=100M$ with 
half-width $\lambda=50$.}\label{table1}
\begin{ruledtabular}
\begin{tabular}{ccccccccc}
$\dot{M}_{\rm{BH}}$&$M\mu$&$A_{0}$&\multicolumn{2}{c}{$ \,\,M\omega $} &\multicolumn{3}{c}{$ \,\,Ms$}&$M_{\rm{BH}}$\\
\cline{4-5}
\cline{6-8}
 &&&1&2&$a$&$b$&$c$ \\

\hline
0.00&0.05&9.42E-05&0.04998&...&3.257E-08&0&0&1.00\\
5E-07&0.05&9.42E-05&0.04993&...&3.272E-08&...&...&1.02\\
5E-06&0.05&9.42E-05&0.04989&...&2.390E-08&2.263E-13&...&1.22\\
5E-05&0.05&9.42E-05&0.04914&0.04951&4.426E-06&-4.755E-10&1.548E-14&7.39\\
1E-04&0.05&9.42E-05&0.04844&0.04908&9.826E-06&-1.787E-09&9.803E-14&54.60\\
\hline
0.00&0.08&6.18E-05&0.07969&...&2.179E-06&0&0&1.00\\
5E-07&0.08&6.18E-05&0.07969&...&2.284E-06&...&...&1.02\\
5E-06&0.08&6.18E-05&0.07959&...&1.886E-06&4.773E-11&...&1.22\\
5E-05&0.08&6.18E-05&0.07905&0.07942&2.571E-05&-4.547E-09&3.318E-13&7.39\\
1E-04&0.08&6.18E-05&0.07825&0.07895&5.309E-05&-1.541E-08&1.693E-12&54.60\\

\hline
0.00&0.10&5.00E-05&0.09945&0.09989 &1.494E-05&0		  &0&1.00\\
5E-07&0.10&5.00E-05&0.09944&0.09988&1.551E-05&...		  &...&1.02     \\
5E-06&0.10&5.00E-05&0.09928&0.09978&1.935E-05&7.147E-11&...&1.22      \\
5E-05&0.10&5.00E-05&0.09878&0.09929&6.962E-05&-1.255E-08&1.455E-12&7.39      \\
1E-04&0.10&5.00E-05&0.09808&0.09873                             &1.054E-04&-2.974E-08&5.194E-12&54.60   \\

\hline
0.00 &0.15&3.37E-05&0.14801&0.14950&3.394E-04&0&0&1.00      \\
5E-07&0.15&3.37E-05&0.14793&0.14954&3.399E-04&...&...&1.02    \\
5E-06&0.15&3.37E-05&0.14778&0.14943&3.450E-04&...&...&1.22      \\
5E-05&0.15&3.37E-05&0.14785&0.14892&3.950E-04&...&...&7.39   \\
1E-04&0.15&3.37E-05&0.14745&0.14807                           &4.360E-04&...&...&54.60\\

\hline
0.00 &0.20&2.54E-5&0.19879&0.19948&6.800E-04&0&0&1.00    \\
5E-07&0.20&2.54E-5&0.19882&0.19955&6.800E-04&...&...&1.02  \\
5E-06&0.20&2.54E-5&0.19868&0.19943&6.830E-04&...&...&1.22  \\
5E-05&0.20&2.54E-5&0.19830&0.19855&7.000E-04&...&...&7.39  \\
1E-04&0.20&2.54E-5&0.19707&0.19787                            &7.197E-04&...&...&54.60\\

\end{tabular}
\end{ruledtabular}
\end{table*}

\subsection{Numerical approach}
\label{sec:numerics}

The solution of the Klein-Gordon evolution equation is computed with the same type of numerical techniques we have 
used in previous works. The reader is addressed in particular to Refs.~\cite{Montero:2012yr,Sanchis-Gual:2015bh} for 
full details on those techniques. As a succinct summary we mention that the evolution equations are integrated using 
the second-order PIRK method introduced in~\cite{Isabel:2012arx,Casas:2014} which 
 allows to handle singular terms that appear due to our choice 
of curvilinear coordinates. The spatial derivatives in the evolution equations are computed using a 
fourth-order centered finite difference approximation on a logarithmic grid except 
for the advection terms for which we adopt a fourth-order upwind scheme.  We also use fourth-order 
Kreiss-Oliger dissipation to avoid high frequency noise appearing near the outer boundary.  Again, we
stress that in the approach adopted in this work, only the scalar field needs to be evolved numerically.
The spacetime is updated following the algebraic equations from Section~\ref{analytic}.

Our simulations employ a logarithmic radial grid, as described in~\cite{Sanchis-Gual:2015sms}. We set the finest radial
resolution close the origin, and a grid spacing of $\Delta r=0.1M$. In the first set of simulations, the outer boundary of the 
computational domain is placed at $r_{\rm{max}}=4\times10^4M$, far enough as to not affect the dynamics of the scalar 
field in the inner 
region during the entire simulation. The time step is chosen to $\Delta t=0.5\Delta r$ which guarantees long-term 
stable simulations. The final time of the numerical evolutions is $4\times10^4M$. In the second set
of simulations, corresponding to the \emph{mirrored} states, 
the radial extension of the numerical domain can be significantly reduced. Specifically we place the outer boundary 
at $r_{\rm{max}}=250M$ which allows us to use ten times less grid points than in the first setup. 
Finally, the time in the second set of simulations extends up to $8\times10^5M$.

\section{Results} 
\label{sec:results}

\subsection{Time evolution of the scalar field amplitude}

\subsubsection{Quasi-bound frequencies}

Quasi-bound states of scalar fields around BHs in the test-field approximation are configurations that have well defined (complex) 
frequencies. The real part gives
the oscillation of the field and the imaginary part gives the rate of decay of the configuration. 
As mentioned in the Introduction, there is a growing body of work which has shown that for some choices of parameters quasi-bound 
states may survive for cosmological timescales around SMBHs and, thus, they are consistent with dark matter models based on 
(ultra-light) scalar fields~\cite{Barranco2011,Barranco:2012qs,Barranco:2013rua,Sanchis-Gual:2015bh,Sanchis-Gual:2015sms}. Such
long-lasting, quasi-bound states have also been found to exist in the nonlinear regime~\cite{Okawa:2014nda}. 

We turn next to discuss the effect of accretion on the quasi-bound states. 
We solve the Klein-Gordon system~(\ref{eq:sist-KG}) using the initial data 
given by Eqs.~(\ref{eq:pulse})-(\ref{eq:iderivatives}) in the rapidly changing gravitational field of an 
accreting BH. As done in~\cite{Sanchis-Gual:2015bh,Sanchis-Gual:2015sms} we analyze the results of 
the simulations by extracting a time series for the scalar field 
amplitude at a set of observation points located at fixed radii 
$r_{\rm{ext}}$ (typically at $r_{\rm{ext}}=100M$). 
We identify the frequencies at which the scalar 
field oscillates by performing a Fast Fourier transform.
It is worth mentioning that
the values of the frequencies do not depend on the location of the observer.

Our main results are summarized in Table \ref{table1}. The first  three columns report our different
choices for the growth rate of the mass of the BH, $\dot{M}_{\rm{BH}}$, the five different scalar
field masses $M\mu$ considered in this work, and the initial amplitude of the Gaussian pulse, $A_{0}$.
The reason we vary the initial amplitude of the pulse for the different values of $\mu$ is to keep the initial scalar 
field energy (almost) constant for the different accretion rates considered, namely $E_{\rm{SF}}^{0}=1.12\times10^{-4}$. 
The oscillation frequencies of 
the scalar field are reported in columns 4 and 5 for the fundamental mode and the first 
overtone, respectively. The imaginary part of the frequencies, shown in columns 6 to 8, is the decay 
rate of the oscillations of the scalar field. From left to right these three columns report the linear, quadratic, 
and cubic parts of analytic fits to numerical data for the different models.
The last column of Table~\ref{table1} also reports the mass of the BH at the end of the simulation, $M_{\rm{BH}}$, 
computed from Eq.~(\ref{eq:bhmass}), whose evolution is displayed in Fig.~\ref{fg:BH}.

\begin{figure}
\begin{center}
\subfigure{\includegraphics[width=0.45\textwidth]{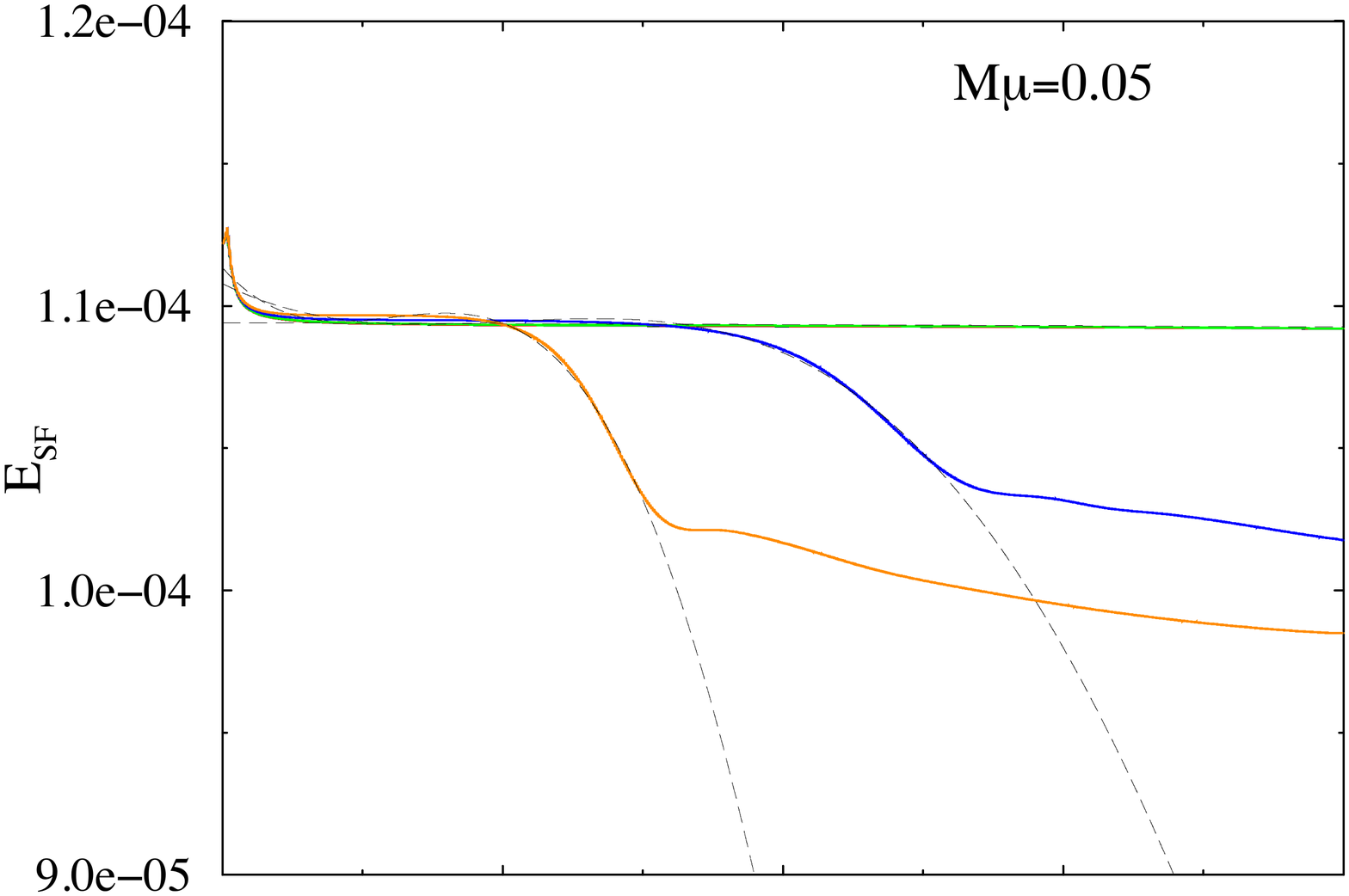}}\vspace{-1.7cm}\\
\subfigure{\includegraphics[width=0.45\textwidth]{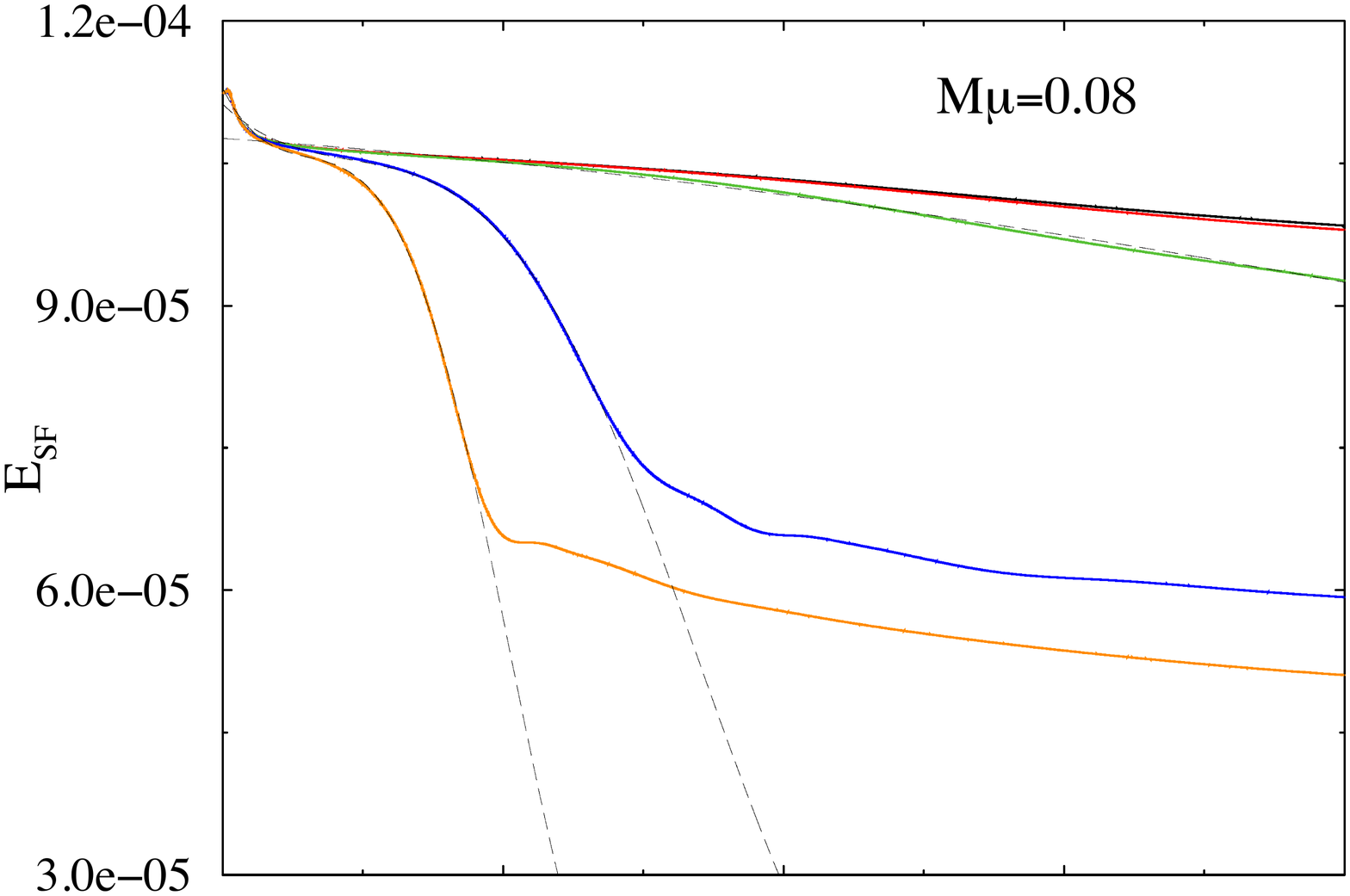}}\vspace{-1.7cm}\\
\subfigure{\includegraphics[width=0.45\textwidth]{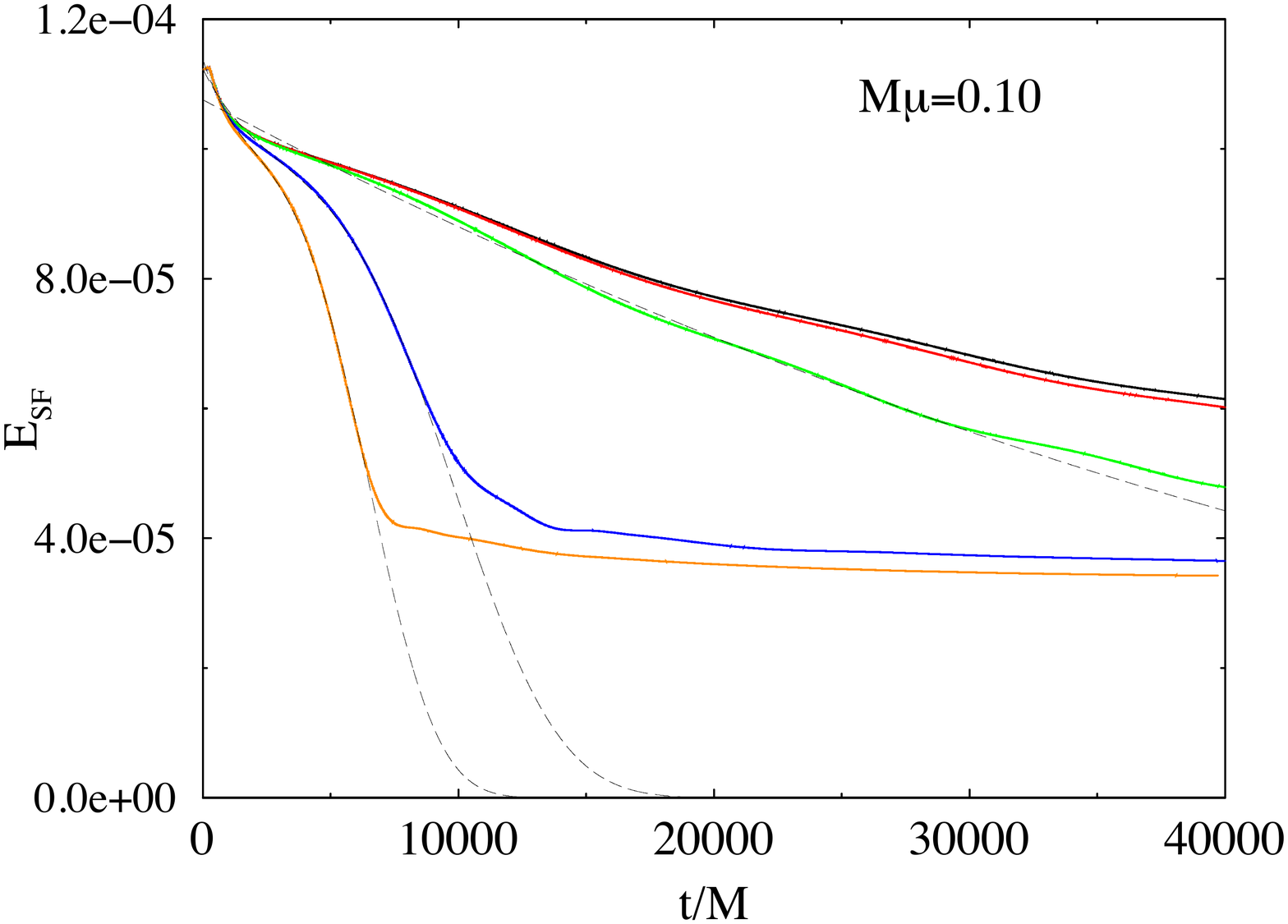}}\vspace{-0.5cm}\\
\caption{Time evolution of the scalar field energy for different values of the BH mass growth rate and for the 
scalar field masses $M\mu=\lbrace0.05,0.08,0.10\rbrace$, from top to bottom. The dashed black lines correspond 
to the exponential fit for the three models with higher BH mass growth rate.}
\label{fg:SF2}
\end{center}
\end{figure}

\begin{figure}
\begin{center}
\subfigure{\includegraphics[width=0.45\textwidth]{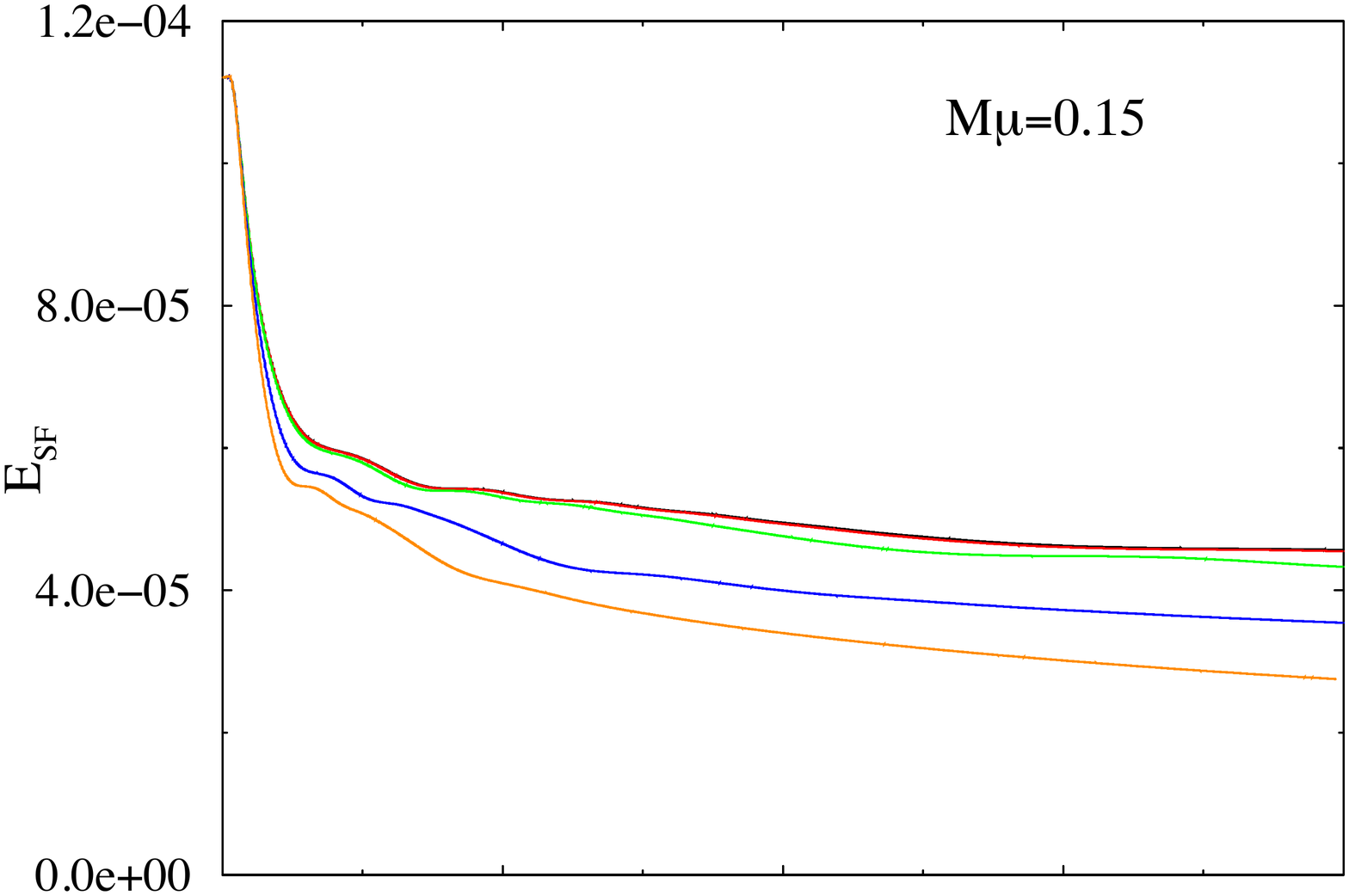}}\vspace{-1.7cm}\\
\subfigure{\includegraphics[width=0.45\textwidth]{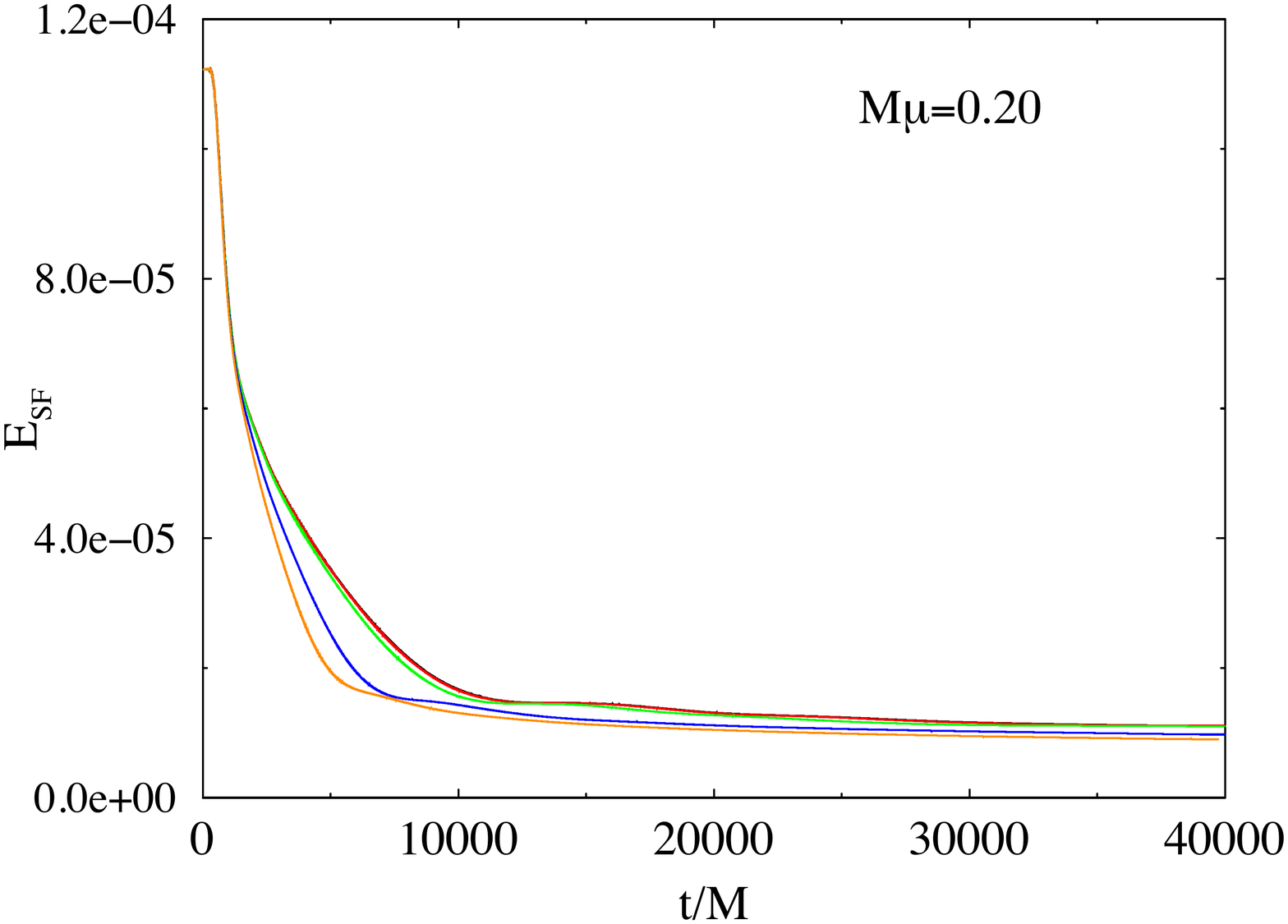}}\vspace{-0.5cm}\\
\caption{Time evolution of the scalar field energy for the masses $M\mu=\lbrace0.15,0.20\rbrace$.}
\label{fg:SF3}
\end{center}
\end{figure}

Figure \ref{fg:SF1} shows the time evolution of the scalar field amplitude and spectra extracted at 
$r_{\rm{ext}}=100M$. In the left panels we plot the evolution for two representative scalar field masses, namely 
$M\mu=0.1$ (top) and $M\mu=0.2$ (bottom), and for the different BH mass growth rates. The criterion for the color of
the different curves and its relation to $\dot{M}_{\rm{BH}}$ follows from Fig.~\ref{fg:BH}. For each panel, the decay 
of the amplitude of the oscillations of the scalar field is seen to depend strongly on $\dot{M}_{\rm{BH}}$, regardless 
of $M\mu$. For the higher growth rates (orange and blue curves), the field amplitudes by the end of the simulations 
have almost practically vanished. Large field amplitudes and long-lasting oscillations are most noticeable for the 
three lower accretion rate models of our sample. The differences for $\dot{M}_{\rm{BH}}=5\times10^{-7}$ with 
respect to the non-accreting case become barely visible only for late times. The right panels of Fig.~\ref{fg:SF1} 
display the corresponding power spectra. They show well-defined oscillation frequencies, best seen in the insets, of 
the quasi-stationary 
states for the models with smaller BH mass growth rate (black, red and green curves), as expected from the bandwidth 
theorem, while the other two models attain wider peaks. 

As expected, in the non-accreting BH case, 
the oscillation frequencies shown in Fig.~\ref{fg:SF1} and reported in Table \ref{table1} are in very good agreement with 
those found in~\cite{Sanchis-Gual:2015bh} in the test-field regime and for the same scalar field masses. Despite in the
current work we are not evolving the spacetime, while we did so in~\cite{Sanchis-Gual:2015bh}, the discarded back-reaction 
on the evolution of the scalar field is negligible because of the small energy of the scalar field in our models.

Our results also indicate that the frequencies of the quasi-bound states decrease as the mass of the BH increases. 
One may infer such behavior from a classical mechanics analogy. Let us consider the solution of the one-dimensional 
wave equation in flat spacetime with moving boundaries. If the boundary moves with a constant velocity $v$ an 
explicit solution can be obtained. Let us consider the oscillation of a string along the $x$ direction. 
The string is constrained at two points, $x=0$ and $x=vt$, and the amplitude of the wave
is given by $h(t,x)$. We are interested in the solution of the equation
\begin{equation}\label{eq:waveflat}
 \partial_{tt} h(t,x) - \partial_{xx}h(t,x) = 0, 
\end{equation}
in the interval $0\leq x \leq vt$, with the boundary conditions,
$h(0,t) = h(vt,t)= 0$
and initial data, $h(x,0) = h_{0}(x)$ $\partial_{t}h(0,x) = u_{0}(x)$.

The general solution of Eq.~(\ref{eq:waveflat}) is a periodic function in $\log [(1-v)t]$ with period $T =\log\left(\frac{1+v}{1-v}\right)$
of the form
\begin{equation}
 h(t,x) = \sum_{n=-\infty}^{+\infty} C_{n} \left\{ \exp \left[i\frac{2n\pi}{T} \tilde\alpha \right]
- \exp\left[i\frac{2n\pi}{T} \tilde \beta \right]  \right\} 
\end{equation}
where $\tilde \alpha = \log (t+x)$ and $\tilde \beta =  \log (t-x)$ and the coefficients $C_{n}$ are determined by the initial 
conditions. What is important to notice in this solution is that the frequency decreases as the velocity of the boundary 
increases. One may think that a similar process happens when the mass of the BH increases since that implies the 
displacement of the position of the horizon and, thus, the displacement of the position of the boundary that confines the 
scalar field.

%

\subsubsection{Scalar field energy}

The time evolution of the energy of the scalar field, computed with Eq.~(\ref{eq:scalar}), is shown in solid lines in Figs.~\ref{fg:SF2} and \ref{fg:SF3} for all scalar field masses, $M\mu=\lbrace0.05,0.08,0.10,0.15,0.20\rbrace$. The dashed lines in the three panels of Fig.~\ref{fg:SF2} correspond to analytic fits of the numerical data, that we discuss below. These two figures show that the initial energy of the scalar field is (almost) the same for all models considered. Both figures show that the decrease of the initial energy
is more significant and faster the larger the scalar field mass and the larger the BH growth. 

Let us first discuss the lightest scalar field models. In figure \ref{fg:SF2}, corresponding to $M\mu=\lbrace0.05,0.08,0.10\rbrace$, 
we see that for the smallest, non-zero value of $\dot{M}_{\rm{BH}}$, the decay of the scalar field energy is very close to the 
non-accreting case (compare red and black curves). For the higher values of the mass flux, the energy at early times begins 
decaying exponentially with a small slope but, at some point, the falling of scalar field energy onto the BH speeds up. The 
appearance of this effect depends on the BH mass growth rate, appearing earlier for higher $\dot{M}_{\rm{BH}}$. After 
some time, the process stops and the scalar field energy settles to an almost constant value, given by the energy of the part 
of the scalar field that has escaped to infinity. There is, however, part of the scalar field still localized around the BH in the 
form of quasi-stationary bound states. 

On the other hand, Fig.~\ref{fg:SF3} shows that for $M\mu=0.15$ and $0.2$,
the decay rate of the energy of the scalar field is much faster than for the lightest models and the energy decreases almost identically 
regardless of the BH mass growth rate. This becomes clear if we check the imaginary part of the frequency reported in column 6 of 
Table \ref{table1} for these scalar field masses. The imaginary part of the frequency does not vary significantly when the growth of the BH mass is faster.

However, as we mentioned before for the smaller values of the scalar field mass displayed in Fig.~\ref{fg:SF2}, 
the energy decay changes from an exponential decay with a linear exponent (the usual feature of a quasi-stationary 
state around a non-accreting Schwarzschild BH) to an exponential decay with both {\it quadratic} and {\it cubic} exponents, 
as shown by the analytic fits. These parts are entirely due to the growth of the BH and are orders of magnitude smaller 
than the linear part of the exponential decay. Therefore, they are only significant for sufficiently long times. 
Columns 6, 7, and 8 of Table \ref{table1} report the values of the rate of decay according to the fit 
$\exp[-c\,t^3-b\,t^{2}-a\,t]$,  where $a$, $b$ and $c$ are the linear, quadratic and cubic coefficients, respectively. 
The dashed black lines in Fig.~\ref{fg:SF2} fit part of the time evolution of the scalar field energy for the three 
models with higher BH growth rate. The values of the cubic, quadratic and linear coefficients increase with the scalar 
field mass, as can be inferred from Table~\ref{table1}.
 
\begin{table*}
\caption{Same as Table \ref{table1} but for the simulation setup corresponding to a scalar field enclosed in a
cavity. The initial Gaussian pulse is, in this case, located at $r_0=50M$ with half-width $\lambda=25$.}\label{table2}
\begin{ruledtabular}
\begin{tabular}{cccccccccccc}
$\dot{M}_{\rm{BH}}$&$M\mu$&\multicolumn{2}{c}{$ \,\,M\omega$}&\multicolumn{6}{c}{$ \,\,Ms$}&$M_{\rm{BH}}$&$t_{\rm{final}}$\\
\cline{3-4}
\cline{5-10}
 &&1&2&$a$&$b$&$c$&$d$&$e$&$f$ \\
\hline
0.00  &0.05&0.05168&0.05781&1.593E-05&0		 &0       	   &0&0&0&1.000&8.0E05      \\
5E-09&0.05&0.05168&0.05781&1.591E-05&9.738E-14 &...               &...&...&...&1.004&8.0E05      \\
5E-08&0.05&0.05167&0.05780&1.590E-05&1.041E-12 &...       	   &...&...&...&1.040&8.0E05     \\
5E-07&0.05&0.05163&0.05776&1.647E-05&7.612E-12 &9.051E-18               &...&...&...&1.221&4.0E05      \\
5E-06&0.05&0.05137&0.05754&5.837E-06&6.122E-10&-5.102E-15&2.179E-20&...&...&3.201&2.4E05      \\
5E-05&0.05&0.05076&0.05688&-2.713E-05&2.762E-08&-1.928E-12&4.847E-17&...&...&6.686&3.8E04      \\
1E-04&0.05&0.05055&0.05655&8.341E-05&4.054E-08&-6.348E-12&3.415E-16&...&...&9.025&2.2E04      \\
\hline
0.00  &0.08&0.08039&0.08431&3.265E-05&0		 &0       	   &0&0&0&1.000&4.0E05      \\
5E-09&0.08&0.08039&0.08431&3.260E-05&2.455E-13 &...             &...&...&...&1.002&4.0E05      \\
5E-08&0.08&0.08038&0.08430&3.272E-05&2.358E-12 &...       	   &...&...&...&1.020&4.0E05     \\
5E-07&0.08&0.08034&0.08426&3.368E-05&2.042E-11 &3.407E-17               &...&...&...&1.221&4.0E05      \\
5E-06&0.08&0.08012&0.08406&5.378E-05&2.742E-10&-3.517E-15&5.500E-20&...&...&2.226&1.6E05      \\
5E-05&0.08&0.07936&0.08337&2.912E-04&-2.237E-08&5.539E-13&5.264E-17&...&...&3.857&2.7E04      \\
1E-04&0.08&0.07981&0.08293&-6.087E-04&5.364E-07&-1.377E-10&1.577E-14&-7.606E-19&1.332E-23&6.050&1.8E04      \\
\hline
0.00  &0.10&0.09975&0.10287&6.355E-05&0		 &0       	      &0&0&0&1.000&4.0E05      \\
5E-09&0.10&0.09975&0.10287&6.348E-05&6.340E-13 &...              &...&...&...&1.002&4.0E05      \\
5E-08&0.10&0.09974&0.10286&6.361E-05&6.300E-12 &...       	      &...&...&...&1.020&4.0E05     \\
5E-07&0.10&0.09969&0.10281&6.898E-05&2.991E-11 &1.283E-16               &...&...&...&1.191&3.5E05      \\
5E-06&0.10&0.09945&0.10262&9.451E-05&1.602E-10&4.284E-15&9.373E-20&...&...&1.733&1.1E05      \\
5E-05&0.10&0.09864&0.10179&6.781E-04&-1.537E-07&1.690E-11&-5.78E-16&7.136E-21&...&3.490&2.5E04      \\
1E-04&0.10&0.09828&0.10151&7.818E-04&-3.295E-07&6.964E-11&-4.692E-15&1.153E-19&...&5.474&1.7E04      \\

\end{tabular}
\end{ruledtabular}
\end{table*}

\subsection{Mirrored states}

\begin{figure}[h!]
\begin{center}
\subfigure{\includegraphics[width=0.44\textwidth]{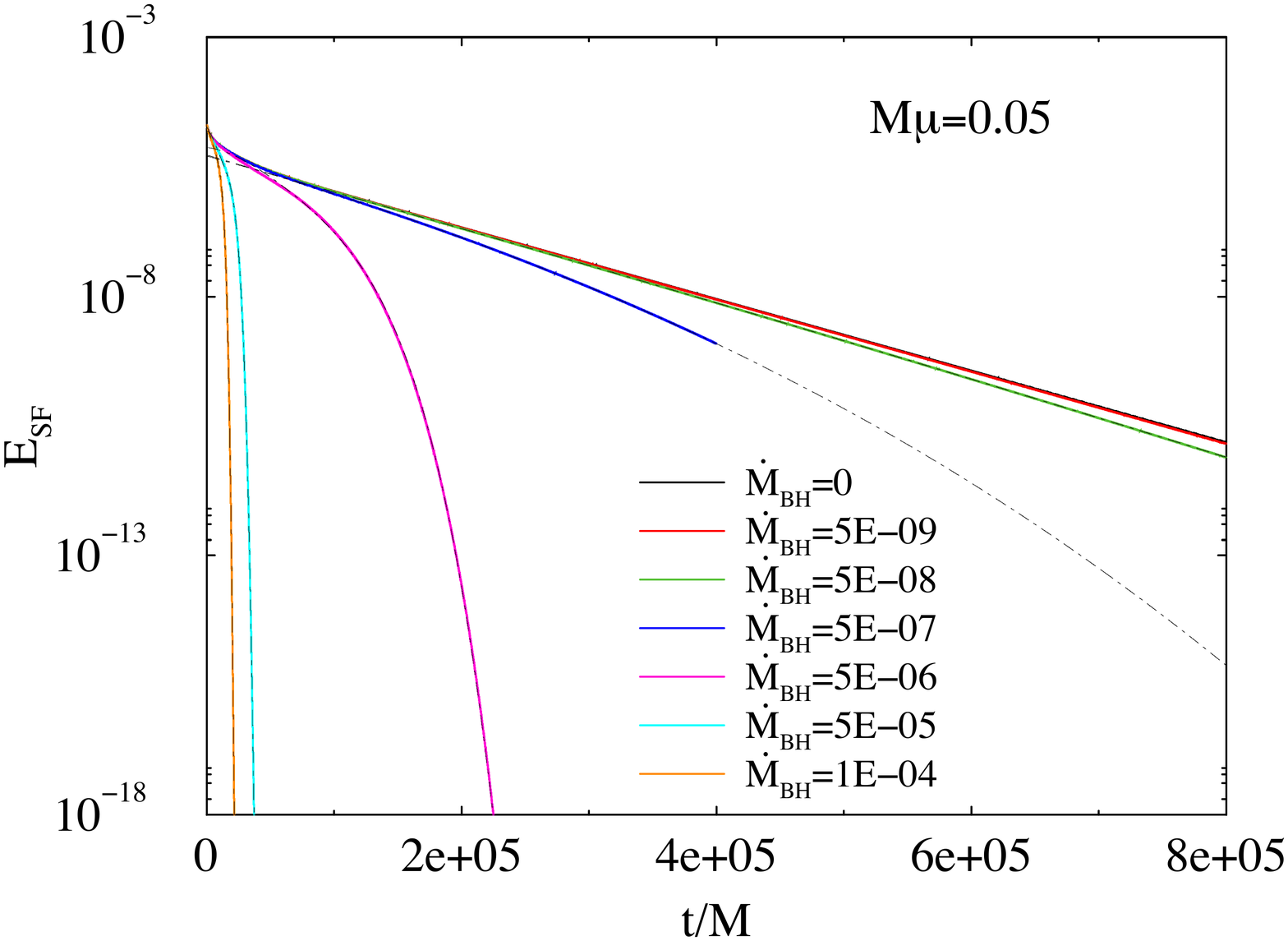}}\vspace{-1.0cm}\\
\subfigure{\includegraphics[width=0.44\textwidth]{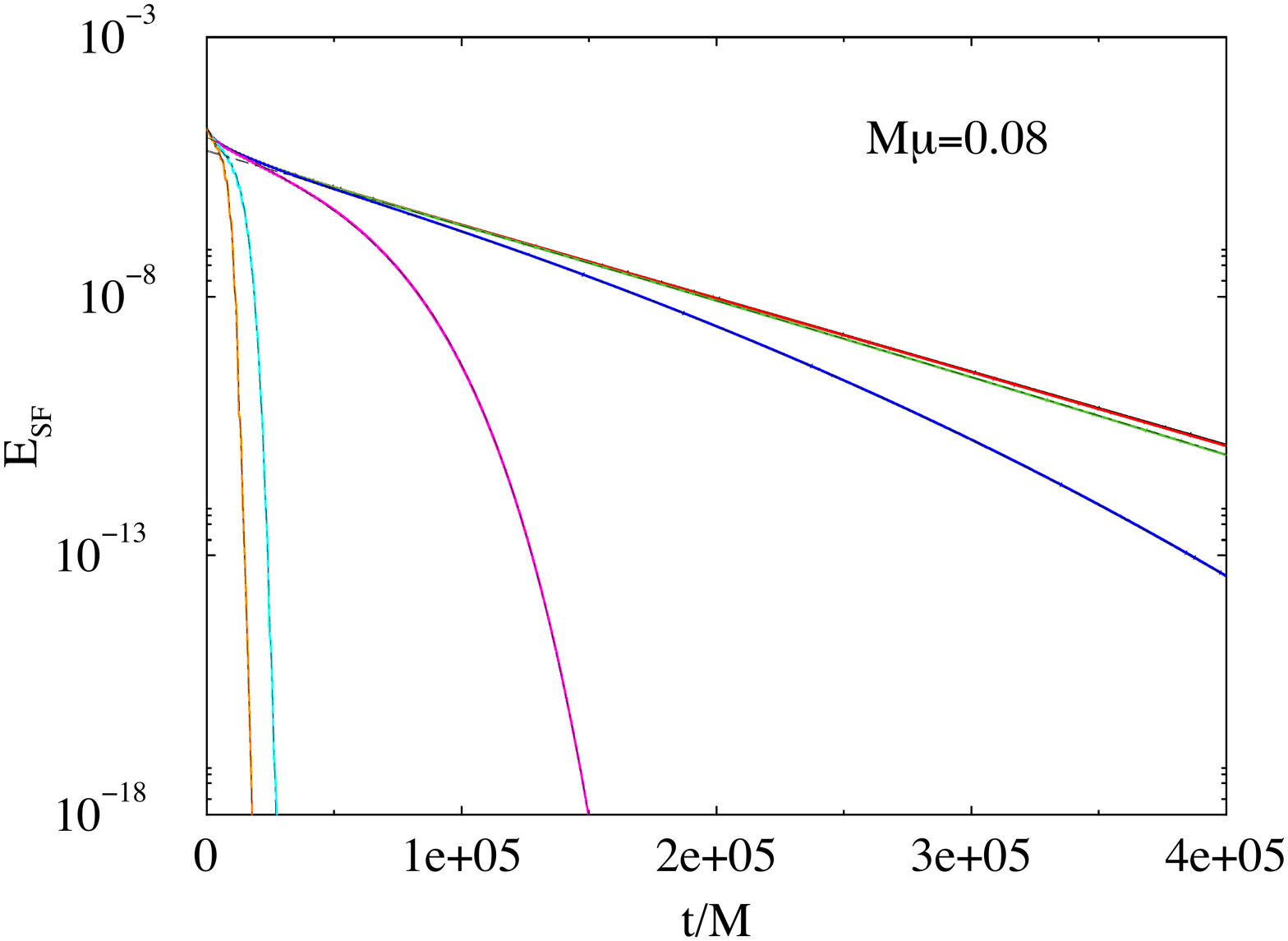}}\vspace{-1.0cm}\\
\subfigure{\includegraphics[width=0.44\textwidth]{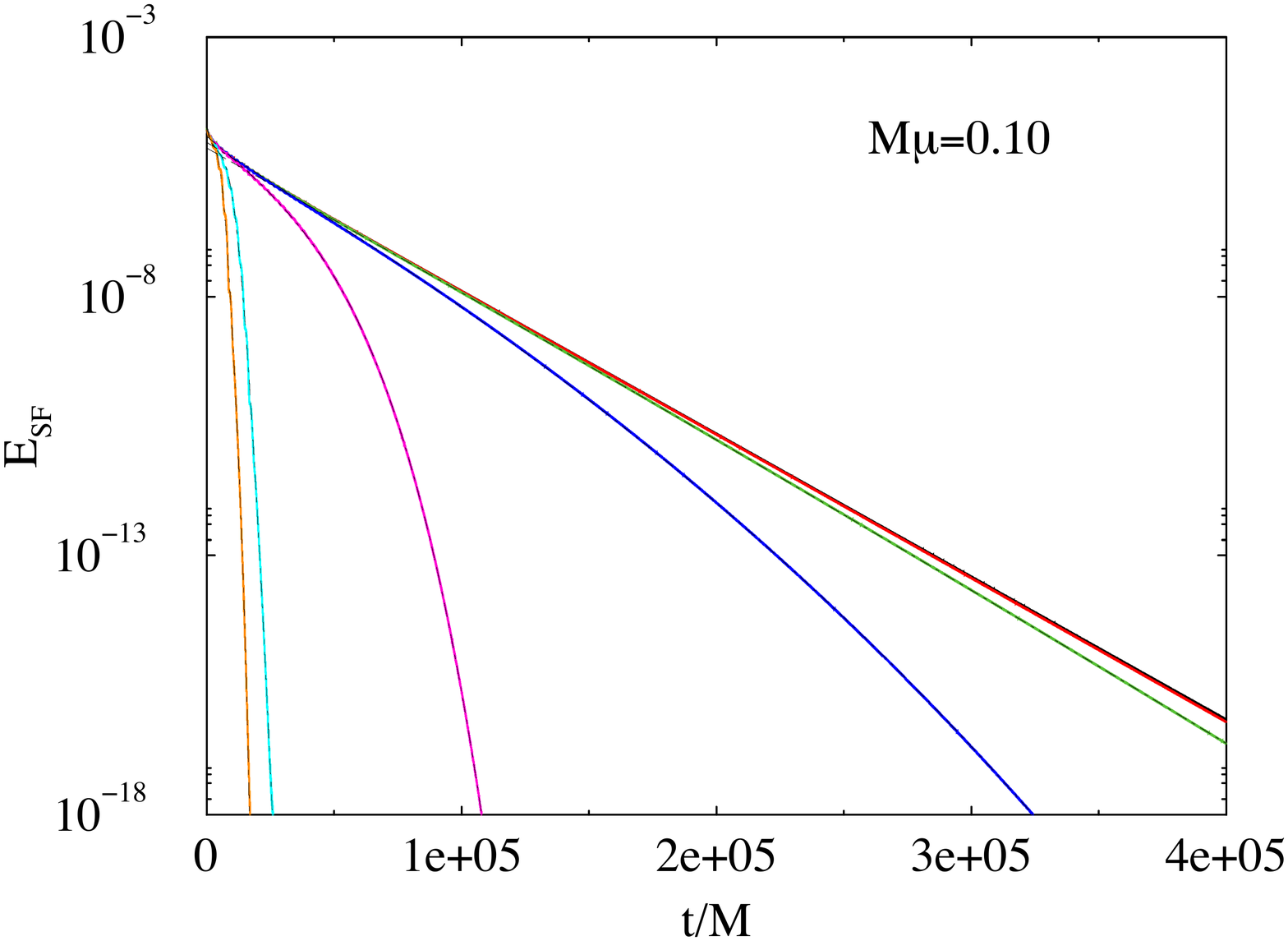}}
\vspace{-0.6cm}
\caption{Time evolutions (solid lines) of the scalar field energy for three scalar field masses $M\mu=\lbrace0.05,0.08,0.10\rbrace$ 
(from top to bottom) and seven different values of $\dot{M}_{\rm{BH}}$ (stated in the legend). The black dashed lines indicate the analytic 
fits. Notice that the scale of the horizontal axis is different in the top panel (the simulation is twice as long) and that the color
criterion is different to that of Fig.~\ref{fg:BH} due to the larger sample of $\dot{M}_{\rm{BH}}$ included in this figure.}
\label{fg:SF6}
\end{center}
\end{figure}

In order to further understand the influence of the growth of the BH mass in the evolution of the scalar field, we turn now 
to describe the results corresponding to the evolution of an additional set of 21 models.
These models correspond to Schwarzschild-like BH spacetimes with 7 different BH mass growth rates, namely 
$\dot{M}_{\rm{BH}}=\lbrace0,\,5\times10^{-9},\,5\times10^{-8},\,5\times10^{-7},\,5\times10^{-6},\,5\times10^{-5},\,10^{-4}\rbrace$ 
and we impose reflecting boundary conditions for the 
scalar field at some radius, as done in \cite{Sanchis-Gual:2016}.  At the location of the mirror, $r=r_{\rm m}$, and beyond, the scalar 
field $\Phi$ is required to vanish. This leads to a discontinuity in the $\Phi$ derivatives. In our simulations the mirror is located 
at $r_{\rm{m}}=200M$ and the boundary conditions for the scalar field are 
\begin{equation}
\begin{split}
\Phi(r_{\rm{m}})&=\Psi(r_{\rm{m}})=\Pi(r_{\rm{m}})=0,\\
\partial_{r}\Phi(r_{\rm{m}})&=\partial_{r}\Psi(r_{\rm{m}})=\partial_{r}\Pi(r_{\rm{m}})=0.
\end{split}
\end{equation}

We consider again a Gaussian distribution for the scalar field, with $r_{0}=50M$ and $\lambda=25$. As mentioned before, using
a mirror is an idealized setup, yet it has pragmatic advantages since it allows us to perform longer runs and to study smaller 
growth rates. The spacetime is still given by the analytic solution
of Section~\ref{analytic} (hence, there are no reflections of the spacetime variables coming in from the outer boundary) and since the scalar field is now 
enclosed in a cavity no part of the scalar field propagates away from the BH. Therefore, all the oscillatory modes will be trapped in this case, not only those with $\omega^2<\mu^2$, and the energy evolution will not be dominated by the asymptotic, scalar field energy minimum. In this idealized situation, the entire scalar field will fall into the BH with time.

The results of this new set of simulations are similar to those discussed in the preceding section. They are 
summarized in Table \ref{table2} for the lightest scalar field models analyzed, namely $M\mu=\lbrace0.05,0.08,0.10\rbrace$. Again, the 
scalar field oscillates with a fundamental frequency and higher overtones (of which Table \ref{table2} only reports the first one), but 
since the higher frequencies decay faster we end up with only the fundamental mode of oscillation. As for the cases without 
mirror, the real part of the frequency decreases with $\dot{M}_{\rm{BH}}$. In the case of a non-accreting BH, the decay of the energy
(and of the amplitude of the oscillations) is exponential with a single linear exponent as expected. As done before, in the case of 
accreting BHs and long evolution times, we fit the energy decay to the analytic formula $\exp[-d\,t^{4}-c\,t^{3}-b\,t^{2}-a\,t]$, 
with linear, quadratic, cubic and quartic terms. These parameters seem to depend on the BH mass growth rate and, probably, we could 
fit it to even higher order 
polynomials if we could reach even longer simulations. Indeed, some of the models reported in Table \ref{table2} can already be 
fitted with higher order terms (even up to 5th or 6th, coefficients $e$ and $f$ respectively) for the current evolution times considered. The 
6th order term is considered for only one particular model, even if it is very small, in order to keep the highest order term always positive and, therefore, drive the fit to tend to zero for $t\rightarrow\infty$.
The coefficients also depend on the  scalar field mass, decreasing for smaller values of $\mu$. 

In Fig.~\ref{fg:SF6} we plot (solid lines) time evolutions of the scalar field energy for the mirrored states of models with 
masses $M\mu=\lbrace0.05,0.08,0.10\rbrace$ (from top to bottom) together with their corresponding fits (black dashed lines). The 
results discussed in the previous section become more clear in this idealized setup. Increasing the BH growth rate speeds up the 
decay of the energy. Moreover, the decay is longer than for the case of quasi-bound states without mirror and, contrary to what happens
when there is an outgoing boundary, the energy 
does not relax to the value corresponding to that part of the scalar field that escapes to infinity. 

\subsection{Comparison with realistic BH mass growth rates}
\label{comparison-observations}

We can estimate the rate of the BH mass growth due to accretion using the bolometric luminosity of an Active Galaxy 
Nuclei (AGN) as~\cite{weihao2003accretion,Tsai:2015wise}
\begin{equation}\label{eq:bolometric}
L_{\rm{bol}}=\frac{\eta \dot{M}_{\rm{BH}}}{(1-\eta)}c^{2}\,,
\end{equation}
where $L_{\rm{bol}}$ is the bolometric luminosity,  $\dot{M}_{\rm{BH}}$ is the BH mass growth rate, $c$ is the speed of light,  
and $\eta$ is the radiative efficiency, for which we take the commonly adopted empirical value of 0.1~\cite{Yu:2002observational}. 
Recently, bolometric luminosities larger than $10^{14}\,L_{\odot}$ have been discovered~\cite{Tsai:2015wise}. According to 
Eq.~(\ref{eq:bolometric}), this corresponds to $\dot{M}_{\rm{BH}}\sim10^{-11}$ in our units.

\begin{figure}[t]
\begin{center}
\subfigure{\includegraphics[width=0.49\textwidth]{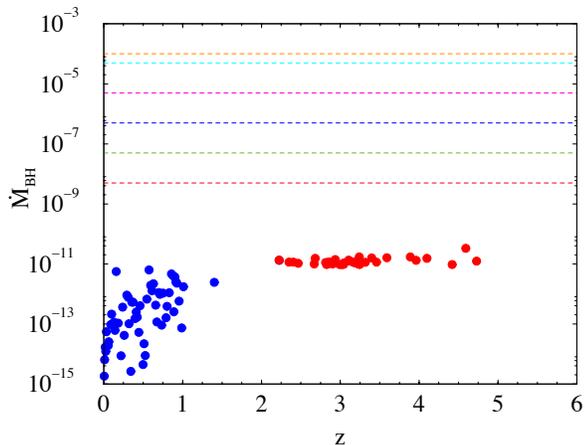}}
\caption{AGN BH growth rates in geometrized units as a function of redshift. The dashed horizontal lines indicate the values
of $\dot{M}_{\rm{BH}}=\lbrace5\times10^{-9}\,(\rm{red}),5\times10^{-8}\,(\rm{green}),5\times10^{-7}\,(\rm{blue}),5\times10^{-6}\,(\rm{magenta}),5\times10^{-5}\,(\rm{cyan}),1\times10^{-4}\,(\rm{orange})\rbrace$ considered in our numerical study. The symbols correspond to observational data 
from~\cite{Tsai:2015wise,woo2002active} (red and blue points, respectively). Our values are about two orders of magnitude larger than the largest observational
value.}
\label{fg:AGN}
\end{center}
\end{figure}

In figure \ref{fg:AGN} we plot several BH growth rates in geometrized units of AGN bolometric luminosities as a function of the 
redshift $z$. We use data from two different samples, namely that of~\cite{woo2002active}, for redshifts between 0.01 and 2.224, and the 
luminosities of WISE-selected galaxies reported recently in~\cite{Tsai:2015wise}, which include redshifts up to $z=4.593$. The larger 
luminosities are obtained precisely for the higher redshifts, corresponding to the growth phase of SMBHs. Fig.~\ref{fg:AGN} shows that 
the maximum growth rate is only about two orders of magnitude smaller than the smallest $\dot{M}_{\rm{BH}}$ we can afford in our
simulations.

The influence of the accretion rate on the energy decay depends on the scalar field mass and decreases for smaller $M\mu$. Therefore, for the expected value of scalar field dark matter models, the actual value of $\dot{M}_{\rm{BH}}$ is expected to be fairly small. However, the growth of SMBH seeds to their actual masses, $\sim10^{9}M_{\odot}$, is believed to take place only during 1\% of their lifetime, i.e.~about $10^{9}$ years. Therefore, the role played by high order terms in the evolution of the energy could become significant during this phase and increase the amount of scalar field that may end up being absorbed by the BH during this period. 

\begin{figure}[t]
\begin{center}
\subfigure{\includegraphics[width=0.49\textwidth]{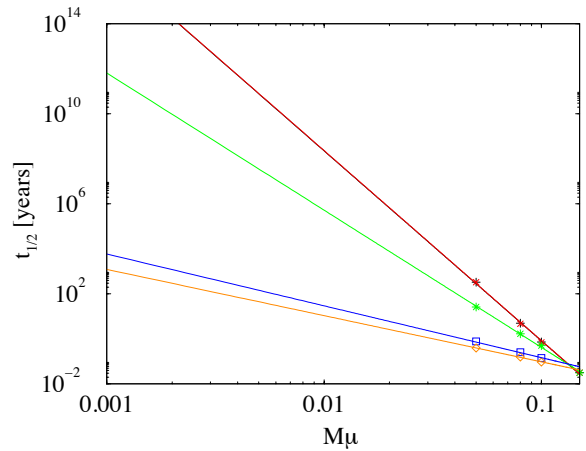}}
\caption{Half-life of the quasiresonant frequencies of the quasi-bound states for different scalar field masses 
and different BH growth rates. The solid lines are least-square fits to the numerical data.}
\label{fg:SF5}
\end{center}
\end{figure}

Finally, we can extrapolate our results to the realistic case of ultra-light scalar field masses. In the case of a test-field
the decay rate of the dynamical resonances is related to the imaginary part of the quasiresonant frequencies, and thus its 
half-life time $t_{1/2}$ is inversely proportional to $\text{Im}(M\omega)$~\cite{Barranco:2012qs}. For an accreting BH and the long 
evolution times reported in this work we have shown that terms higher than linear are important to capture the decay
rate of the energy of the scalar field. We can then use the relation $E_{\rm SF}=E^0_{\rm SF}\exp{[-(at+bt^2+ct^3)]}$ to solve
for $t_{1/2}$ which corresponds to the time when the energy of the scalar field has decreased to half its initial value. The
result of this exercise is depicted in Fig.~\ref{fg:SF5}, which shows $t_{1/2}$ as a function of $M\mu$ for all five values of
$\dot{M}_{\rm{BH}}$ used in the ``unmirrored" setup simulations. (We use in this figure the same color criterion for $\dot{M}_{\rm{BH}}$ as
defined in Fig.~\ref{fg:BH}.) The symbols in the figure indicate the values of  $t_{1/2}$ for the lighter values of the scalar 
field mass ($M\mu=0.05, 0.08$ and 0.1)
used in our simulations. Following~\cite{Barranco:2012qs} we do a least squares fit to these data and extrapolate our results
to lower values of $M\mu$. Fig.~\ref{fg:SF5} shows that even for high mass accretion rates, e.g.~$\dot{M}_{\rm{BH}}=5\times 10^{-7}$ 
(red curve) quasi-bound states around accreting BHs can survive for timescales significantly longer than the age of the Universe.

\section{Conclusions}
\label{sec:summary}

We have studied the properties of scalar field quasi-bound states in the background of an accreting, spherically symmetric BH. 
We have assumed that the BH mass grows due to matter accretion, describing the effect of the increase of the BH mass on the 
properties of the quasi-bound states of the surrounding scalar field distribution. For our study we have solved the 
Klein-Gordon equation numerically, mimicking the evolution of the spacetime through a sequence of exact Schwarzschild BH solutions of 
increasing mass, using analytic expressions which depend only on the BH mass parameter. 
Our numerical approach has been limited to spherical symmetry and has been based on spherical 
coordinates and a PIRK numerical scheme for the time update of the Klein-Gordon equation. To study the effect of accretion on 
the scalar field evolution in affordable computational times, we have resorted to large growth rates, at best two orders of magnitude 
larger than the  largest observational estimations. 

By performing a Fourier transform of the time series of our numerical data we have been able to characterize the scalar field 
states by their distinctive oscillation frequencies. It has been found that the frequencies decrease with increasing BH mass. 
Moreover, accretion results in a significative increase of the exponential decay of the scalar field energy due to 
the presence of terms of order higher than linear in the exponent. These terms 
are zero in the non-accreting Schwarzschild BH case, resulting in a linear exponential fit. 
Our computational setup has considered both outgoing and reflecting boundary 
conditions at the outer radial boundary, the latter describing a scalar field enclosed in a cavity.
By imposing reflecting boundary conditions at a finite distance the scalar field does not escape to infinity and we can isolate the 
influence of accretion. Such configuration, albeit artificial, has helped us to study the higher order terms that appear in the exponential decay 
of the energy. Finally, we have compared our BH mass growth rates with 
estimates from observational surveys, and we have been able to extrapolate our results to realistic values of the scalar field mass 
$M\mu$. We have found that even for the high mass accretion rates considered in this work, e.g.~$\dot{M}_{\rm{BH}}=
5\times 10^{-7}$, quasi-bound states around accreting BHs can survive for cosmological timescales. The results obtained in 
this paper add further support to the intriguing possibility of the existence of dark matter halos based on (ultra-light) scalar 
fields surrounding SMBHs at galactic centers.

\begin{acknowledgments}
This work was supported in part by the Spanish MINECO (AYA2013-40979-P), 
by the Generalitat Valenciana (PROMETEOII-2014-069), by the 
CONACyT-M\'exico, ICF-UNAM, and by the Max-Planck-Institut f{\"u}r Astrophysik. The computations have been 
performed at the Servei d'Inform\`atica de la Universitat de Val\`encia.
 \end{acknowledgments}

\bibliography{num-rel}

\end{document}